\documentclass{rspublic}

\usepackage[dvips]{color}       
\usepackage{graphics}           
\usepackage{tikz}               
\usetikzlibrary{arrows,scopes}  
\usepackage{bm}                 
\usepackage{amsmath}            
\usepackage{amsfonts}
\usepackage{amssymb}
\usepackage{latexsym}           

\usepackage[round,colon,authoryear]{natbib}
\bibpunct[, ]{(}{)}{,}{a}{}{,}
\newcommand{\identity}{\leavevmode\hbox{\small1\kern-3.2pt\normalsize1}}

\begin{document}

\title
      [Quantum Analogue Computing]
      {Quantum Analogue Computing}

\author
       [V.~Kendon, K.~Nemoto, W.~J.~Munro]
       {Vivien M.~Kendon$^1$, Kae Nemoto$^2$ and William J.~Munro$^{2,1}$}

\affiliation{$^1$School of Physics and Astronomy,
	University of Leeds, Leeds LS2 9JT, UK\\
	     $^2$National Institute of Informatics, 2-1-2 Hitotsubashi,
	Chiyoda-ku, Tokyo, 101-8430, Japan}	

\label{firstpage}

\maketitle

\begin{abstract}{quantum information, quantum computation, quantum simulation}
We briefly review what a quantum computer is, what it promises to do for
us, and why it is so hard to build one.  Among the first applications
anticipated to bear fruit is quantum simulation of quantum systems.
While most quantum computation is an extension of classical
digital computation, quantum simulation differs fundamentally in how the
data is encoded in the quantum computer.
To perform a quantum simulation, the Hilbert space of the
system to be simulated is mapped directly onto the Hilbert space of the
(logical) qubits in the quantum computer.  This type of direct
correspondence is how data is encoded in a classical analogue computer.
There is no binary encoding, and increasing precision becomes exponentially
costly: an extra bit of precision doubles the size of the computer.
This has important consequences for both the precision and error
correction requirements of quantum simulation, and significant open questions
remain about its practicality.
It also means that the quantum version of analogue
computers, continuous variable quantum computers (CVQC) becomes an equally
efficient architecture for quantum simulation.
Lessons from past use of classical analogue computers can help
us to build better quantum simulators in future.
\end{abstract}

\section{Introduction}
\label{sec:intro}

It is a quarter of a century since the seminal work of \cite{feynman82a}
and \cite{deutsch85a} introduced the concept of quantum computation,
and fifteen years since \cite{shor95a} produced an efficient quantum
algorithm for factoring large numbers, thereby raising the game for
cryptographic schemes by suggesting a quantum computer could break
them more easily.
Yet working quantum computers are still just toys in a test tube performing 
calculations a child can do in their head.  This is not for want of trying.
Spurred on by the seminal work of \cite{shor95b,laflamme96a,steane96a},
who showed how quantum error correction can be used to protect quantum
systems from decoherence for long enough to run a computation,
the growth in research has been phenomenal on both theoretical
and experimental aspects of the challenge to construct a quantum computer.
Steady progress has been made over the intervening
years, and there are a plethora of promising architectures on the
drawing board and the laboratory bench.
There is currently no clear front runner in the pack.  We are at a stage
of development equivalent to that of classical computers before
the advent of silicon chips (a mere fifty years ago).
We know a lot about what a quantum computer
should be able to do, and what components it needs, but we haven't
identified the best materials to build it with,
nor what it will be most useful for calculating.


\section{Quantum computing}
\label{sec:qc}

\begin{figure}
\begin{center}
\resizebox{0.85\columnwidth}{!}{\rotatebox{0}{\includegraphics{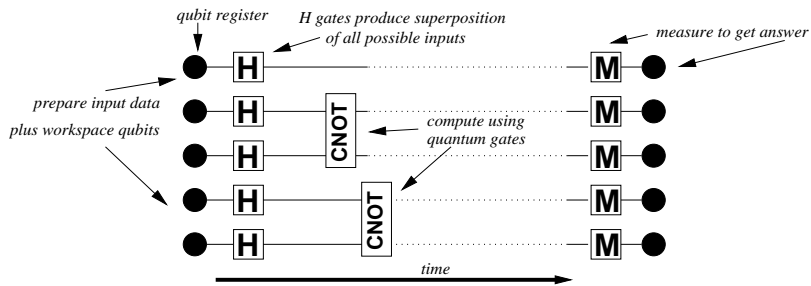}}}
\end{center}
\caption{Digital quantum computation (gate model).}
\label{fig:qcomp}
\end{figure}

\citeauthor{feynman82a} and \citeauthor{deutsch85a} both (independently)
perceived that a superposition of multiple quantum trajectories looks like
a classical parallel computer, which calculates the result
of many different input values in the time it takes for one processor
to do one input value.  Except the quantum system doesn't need a stack
of CPUs, the parallel processing comes `for free' with 
quantum dynamics, potentially providing a vast economy of scale over
classical computation.  Of course, there is no `free lunch'.
Despite doing all the calculations in superposition, we can only 
access one of the results when we measure the final state of the
quantum computer.  Figure \ref{fig:qcomp} illustrates schematically
how a simple gate model quantum computer works.
%

Programming for a quantum computer is thus trickier
than classical programming: you have to include a step to 
select the answer you want (at least with high probability) out of all
the possible results.
Quantum programming techniques can be grouped into a few distinct
types.  Promise problems, like \cite{deutsch92a},
hidden subgroup problems, like \cite{shor95a} for factoring,
search algorithms like \cite{grover96a}, and quantum versions
of random walks (see \cite{ambainis03a} for a review, and
\cite{kendon06a} for a gentle introduction).
Some techniques are best understood as a different architecture
rather than a different programming method.
Minimisation problems, such as finding ground state energies,
or satisfiability problems (kSAT) are best tackled using
adiabatic quantum computation \citep{farhi00a}.
Adiabatic quantum computation has been proved to be equivalent
to the standard gate model of quantum computing \citep{aharonov04a},
meaning that we will be able to run adiabatic quantum algorithms
on any quantum computer.

Many of these potential applications will require very large quantum
computers before they gain an advantage over classical computers.
For example, the best classical factoring to date
\citep{bahr05a} can handle a 200 digit
number (RSA-200), which is about 665 bits.  For an $n$-bit input number,
a basic implementation of
Shor's quantum factoring algorithm needs $2n$ qubits in the QFT register
plus $5n$ qubits for modular exponentiation = $7n$ logical qubits, so a
665 bit number needs 4655 logical qubits.
For a quantum computer of this size, error correction will be essential.
How much this increases the number of physical qubits
depends on the error rates.
The closer to the threshold rates of $10^{-3}$ to $10^{-4}$, the more
error correction is necessary \citep{devitt09a}.
These threshold error rates are smaller than any experiment has yet achieved.
For low error rates, maybe 20--200 physical qubits per logical qubit are
sufficient, while for high error rates (close the threshold) it blows up
quickly to around $10^5$ per logical qubit.
This suggests we may need Teraqubit quantum computers to break factoring.
While the scaling favours quantum, the crossover point is high.

The picture is rather more promising for using a quantum computer to 
simulate a quantum system, \citeauthor{feynman82a}'s original inspiration
for quantum computation.
For example, a system consisting of $N$ 2-state quantum particles 
has $2^N$ possible states,
and in general, it could be in superposition of all of them.
Classical simulation of such a system requires
one complex number per state, which amounts to $2^{N+1}\times$size-of-double.
With 1 Gbyte of memory, a system of $N=26$ of the 2-state quantum
particles can be accommodated.
The record set by \cite{deraedt06a} is $N=36$ in 1Terabyte of memory.
Each additional particle \textit{doubles} the amount of memory required.
Simulating more than 40 or so qubits in a general superposition state
is beyond current classical capabilities.  If the system is restricted
to only part of its full Hilbert space, not all superpositions
are needed, and with appropriately designed methods to take this
into account, larger classical simulations are possible
\citep{verstraete04a}.
This suggests that quantum computers with upwards of 40 logical qubits could
perhaps do something useful for us we can't accomplish with classical
simulation, in stark contrast to the Teraqubits required for factoring.

While a classical simulation allows the full quantum state to be tracked 
throughout, data extraction from quantum simulations is not nearly so easy.
Each run provides only a single snapshot of the outcome of a measurement
on the superposition state of the system.
But methods have been designed for extracting key parameters, such as
energy gaps \citep{wu02a},
or correlation functions \citep{somma02a}
from quantum simulations.
Open problems in the theory of nanoscale superconducting materials
are one example where quantum simulation could make an impact on
the development of the field.

Quantum simulation has been demonstrated \citep{somaroo99a},
for small systems using NMR quantum computers.  
So what is stopping us from building a quantum computer with 40 to
100 qubits tomorrow?  Quite simply, this is extremely demanding technology
to design and build.  It requires single quantum particles to be
localised and controlled for long enough to apply hundreds or thousands of
precise manipulations to them, singly or in pairs, and then measured
individually with high fidelity.  And throughout the process, 
unwanted interactions from nearby material or light must be prevented, or
minimised and any remaining effects reversed using error correction
techniques.  Current state of the art for a quantum simulator is $\lesssim 10$
qubits.  Scaling up to 50 with a few hundred manipulations is believable with
the steady rate of development the experimentalists have been 
delivering.
And it is believable that at around 50 qubits, a quantum simulator
could perform a short computation without needing significant
amounts of error correction.  Going beyond this without error
correction is stretching credibility based on what we know today.
But error correction designed for digital quantum computers
doesn't work well enough for a quantum simulator.  We will elaborate
on this point in section \ref{sec:precision}, after discussing
simulation in general, to set the problem in context.

\section{Simulation}
\label{sec:sim}

Computer simulations play an important role in most areas of science.
Numerical methods kick in wherever analytic calculations become
too complicated or intractable, which can happen even for simple systems,
like three bodies moving under their mutual gravitational interactions.
We use computer simulation to test our models
of the real world, by calculating in detail what they predict,
then comparing this with our experimental observations.
We don't expect perfect agreement, both experiments and computations
have limited precision, and we may know the model is only an approximation
(to simplify the computation). 
If our calculations and observations agree as well as we anticipate, this
is evidence our models are good, and hence we understand at some level
how the system works.
We may then use our computer simulations to predict things we haven't yet
observed, or provide more details of processes that are hard to observe
by experiment.
We may even simulate systems that can't, or don't, exist in nature,
to better understand why nature is the way it is.

That computation of any sort works in a useful way is not trivial. 
An accurate calculation of the trajectory of a space probe
that takes five years to reach its destination is crucial: we cannot
just launch many test probes and see where they end up in order to
work out which trajectory we want.
One key difference between a classical
physical system and a computer simulation of it is that we represent
it using binary numbers (0 and 1) in a register in the computer.
This is like the difference between counting on your fingers and 
writing down the number 8.  Eight has just one symbol in place of your 
eight fingers.  Eight people have 64 fingers between them and that only 
took two symbols to write down.  This makes for a huge saving in the amount 
of memory a computer needs compared to the size of the physical system,
see figure \ref{fig:unary}.
\begin{figure}
\begin{center}
\begin{minipage}{0.48\textwidth}
\begin{tabular*}{\textwidth}{r@{\extracolsep{\fill}}r@{}r}
Number & Unary & Binary\\
0 &    & 0 \\
1 & $\bullet$ & 1 \\
2 & $\bullet\bullet$ & 10\\
3 & $\bullet\bullet\bullet$ & 11\\
4 & $\bullet\bullet\bullet\bullet$ & 100\\
$\cdots$ & $\cdots$ & $\cdots$ \\
$N$ & $N \times \bullet$ & $\log_2N$ bits \\
\end{tabular*}\\
\end{minipage}
\end{center}
\caption{Binary counting needs exponentially less memory than unary.}
\label{fig:unary}
\end{figure}
Binary numbers mean counting using base two,
whereas we usually use base ten.  The saving 
in memory is huge whichever base we use, but the inner workings of the 
computer are easier to arrange using binary.

A second important reason why computer simulations work depends on
the complexity of the model compared with the complexity of the
system being studied.  Stars and planets are highly complex
systems, but we don't need to model most of the details to predict to
very high precision how the planet will orbit the star.  Gravitation
is a supremely simple theory compared with systems it can 
correctly describe the behaviour of.  This is part of the
way our world appears to be, described as the ``unreasonable effectiveness of
mathematics'' by \cite{wigner62a}.
More discussion of simulating complex classical systems,
and how to apply quantum computers to this task,
can be found in \cite{harris10a}, here we are specifically
interested in simulating quantum systems.

\section{Efficiency and precision}
\label{sec:precision}

For practical purposes, an efficient computation is one that
gives answers in relatively short human time scales (seconds,
minutes, hours or days) at a cost we can afford to pay.
Computer scientists make quantitative comparisons
between different algorithms by associating a cost with each
step of the algorithm and with the amount of memory required to
hold the data.  This abstracts the idea that physical computers
have a limited size (memory) and maximum
rate of elementary calculation steps per unit time.
In most cases, `efficient' means the algorithms use resources
that scale as a simple polynomial
(like $N^2$) of the size $N$ of the problem.

If we could efficiently simulate a quantum 
algorithm on a classical computer, we would
immediately have a classical algorithm that is sufficiently
good to render the quantum algorithm superfluous.
To prove an algorithmic speed up for a quantum algorithm,
we have to prove that no classical algorithm of any type can do as well:
such proofs are in general very difficult.
Shor's algorithm \citep{shor95a} is only the `best known' algorithm
for factoring, there is no proof something classical and
faster can't be found in future.

Just as we can't usually make a perfect measurement that determines
exactly how long or heavy something is, we also generally have to
represent numbers in a computer to a fixed precision.  As the computation
proceeds, the errors due to the fixed precision slowly grow.  For very
long computations, they can overwhelm the answer and render it useless.
The accuracy we require for our answer thus places another demand on 
how much memory and time we need to perform our computation.
Starting out with a higher precision to represent our numbers allows
high precision for the answer.  If we represent our numbers as floating
point with an exponent, e.g., $0.1011101101 \times 2^{4}$, then for an
error of size $\varepsilon$, the number of bits (binary digits) of
precision is $\log_2(1/\varepsilon)$.
For the example just given, which has ten bits of precision,
$\varepsilon=2^{-10}$.
Efficient algorithms require resources (time and memory) that scale
as $\log_2(1/\varepsilon)$.   Algorithms running on digital computers,
that store their numbers in this binary format, usually have this
logarithmic scaling for precision.
However, for computers that store information in a unary encoding,
such as quantum simulators, the errors will scale as $1/\varepsilon$,
thus requiring exponentially more resources for the same increase
in precision as their binary encoded counterparts.

Approximating real numbers to a fixed precision introduces errors into
calculations that grow larger as the computation proceeds.
If the computation requires many steps, it may be necessary
to use a higher precision to avoid the errors swamping the final answer.
As well as errors arising from the limited precision,
the operation of the computer itself may be imperfect and
introduce random errors.
In our present day classical computers, such errors are so
unlikely to happen we can forget about them.
In contrast, error correction for digital quantum computation will be 
essential beyond about 100 qubits and 100 computational steps.

\section{Quantum simulation of quantum systems}
\label{sec:qsim}

The main reason why classical simulations of quantum systems
are not efficient is because the memory required scales 
exponentially in the system size, due to the exponential
number of quantum superpositions.  This can be solved by
using the efficient storage achieved by quantum computers.
Mapping the Hilbert space of the system directly
onto the Hilbert space of the quantum computer 
gives us efficient memory use.   We also require
that a quantum simulator can run using an efficient number of time
steps.  \cite{lloyd96a} showed that the Hamiltonian evolution
of the quantum system can be decomposed into a sequence of
standard Hamiltonians by using the Trotter
approximation.  Each standard Hamiltonian is applied
for a short time to build up the desired evolution.
The number of steps required scales polynomially in the accuracy,
so this part of the algorithm is also efficient.

However, any small errors or imperfections in the operation of
the quantum simulator will affect the result in a linear rather
than logarithmic way, because the system is mapped directly onto
the quantum simulator, rather than being binary encoded into numbers
in qubit registers.  Consequently, we will need exponentially
more resources to correct for these errors.  This raises the
prospect of crippling error correction requirements for 
practical quantum simulators \citep{brown06a}.

Quantum simulation has much in common with analogue computation in the way 
the data is encoded.  It shares the problem that an extra bit of
precision requires a doubling of the resources.
However, this suggests two avenues for 
the development of quantum simulation that have not yet been well-explored.
Firstly, looking back to the era of analogue computation can teach
us lessons for how to make quantum simulation practically useful,
despite the unfavorable scaling of the resources.
Secondly, we can consider quantum versions of analogue computation for
quantum simulation architectures.  We discuss both these ideas in
the next sections.

\section{Analogue computing}
\label{sec:analogue}

Despite the exponential saving in memory gained through binary encoding,
the earliest computers didn't use it.  From the invention of the 
astrolabe 
for plotting the heavens in around 200 BC,
through the slide rule, and mechanical differential analyser 
\citep{kelvin1876a}, these computational devices represented the quantity
they were computing as the size of some part of the apparatus.
If you wanted a more accurate answer, you had to build 
a larger device: for another bit or precision, it 
would need to be twice the size.  Yet, for the problems these
machines were designed to solve, they were very effective in practice.

\begin{figure}
\centering
\begin{tikzpicture}
\tikzstyle{block} = [draw,fill=black!15,minimum size=2em]
\tikzstyle{branch}=[fill,shape=circle,minimum size=3pt,inner sep=0pt]

\node at (0,0.25) (x1) {$x$};
\node at (0,-0.25) (y1) {$y$};
\draw[-] (x1) -- (1,0.25);
\draw[-] (y1) -- (1,-0.25);
\node[block] at (1,0) (block1) {$+$};
\node (result1) at (2.5,0) {$x+y$} edge [-] (block1);
\node (adder) at (1,0.7) {\textsc{adder}};

\node at (3.5,0.0) (x2) {$x$};
\draw[-] (x2) -- (4.5,0);
\node[block] at (4.5,0) (block2) {$a$};
\node (result2) at (6,0) {$a\cdot x$} edge [-] (block2);
\node (multiplier) at (4.7,0.7) {\textsc{multiplier}};

\node at (7,0.25) (x3) {$x$};
\node at (7,-0.25) (y3) {$y$};
\draw[-] (x3) -- (8,0.25);
\draw[-] (y3) -- (8,-0.25);
\node[block] at (8,0) (block3) {$\int$};
\node (result3) at (9.5,0) {$\int x \mathrm{d}y$} edge [-] (block3);
\node (integrator) at (8.3,0.7) {\textsc{integrator}};

\node (x4) at (10.5,0) {};
\node (result4) at (12,0) {$x$} edge [*-|] (x4);
\node (constant) at (11.4,0.7) {\textsc{constant}};

\end{tikzpicture}
\caption{GPAC operation set.}
\label{fig:gpac}
\end{figure}
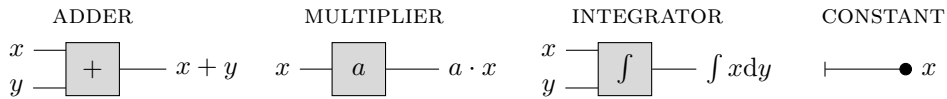

\cite{shannon41a} provided a theoretical model,
the General Purpose Analogue Computer (GPAC),
which is a mathematical description of the differential analyser.
The GPAC consists of a set of nonlinear boxes, connected by their
inputs and outputs.  
Four basic operations are sufficient, shown in figure \ref{fig:gpac}.
There are a few rules for how the boxes can be connected.  
Each output of a box goes to at most one input;
inputs cannot be interconnected, i.e., there is no ``splitter''
to allow the same input to go to more than one box;
similarly, outputs cannot be interconnected.
It has been proved by \cite{pourel78a}, and \cite{lipshitz87a}
that the set of functions generated this way
is the set of differentially algebraic functions.
In other words, we can solve any ordinary differential equation with
this type of device, given appropriate boundary conditions as part of
the input.
This abstract notion of computation can be realised in many different
physical systems, such as water in pipes or electronic circuits.
Such systems were quite common in university science departments
before digital computers became widespread.

There have been a number of suggestions for how
to expand the model by adding more operations to the available set
e.g., \cite{rubel93a,silvagraca04a,mills08a}.
These are mathematical extensions and it isn't clear whether they can
be physically constructed, raising the question of just exactly 
what we can in practice compute given the way the physical world works.
\cite{rubel81a} observed that there are simple differential equations
that are universal in the sense that they can approximate any other
differential equation to arbitrary accuracy.
In practice, these probably aren't useful
for computation, attempts to build a circuit that implements such a
differential equation turn out to be completely unstable \citep{mills09a}.
While \cite{rubel89a} proved that a classical digital computer can
efficiently simulate an analogue computer, the reverse question of
whether an analogue computer can simulate a digital computer
efficiently is open.

Analogue computers might still provide a useful function in our digital 
world: for certain calculations they are extremely fast.  When a specific
task is required repeatedly (e.g., video rendering in real time),
a dedicated analogue ``chip'' might outperform a digital circuit.

\section{Continuous variable quantum computing}
\label{sec:cvqc}

There is a fully quantum version of analogue computing, 
usually known as continuous variable quantum computing (CVQC) and
first described by \cite{lloyd99b}.
The information is encoded in the eigenstates of a 
continuous-spectrum operator such as position $\hat{x}$ or
momentum $\hat{p}$.  Computations are carried out by
manipulating the physical states.
Position and momentum are a conjugate pair of operators,
which are orthogonal in the sense that $[\hat{x},\hat{p}] = i$ up to
a real normalisation constant.
They act on the Hilbert Space $L_2 (\mathbb{R})$,
the space of square-integrable functions over $\mathbb{R}$
(square-integrable is important as it corresponds to being normalisable).
To perform a quantum computation, 
we create an initial state, evolve the state in a prescribed way
by applying an appropriate Hamiltonian to the state, and then 
perform a measurement from which we can extract the results.

Universal computation can be done in CVQC using a small set of elementary
operations.  Any Hamiltonian can be written as a Hermitian polynomial in the 
position and momentum operators $\hat{x}$ and $\hat{p}$, and we can
generate any Hermitian polynomial in $\hat{x}$ and $\hat{p}$ by
using the following set of operators \citep{lloyd99b}:
\begin{enumerate}
	\item simple linear operations, $\{\pm\hat{x},\pm\hat{p}\}$,
	and quadratic, e.g., $\hat{x}^2 + \hat{p}^2$,
	\item a non-linear operation at least cubic i.e., $\hat{x}^3$
	or $\hat{p}^3$, the Kerr Hamiltonian
	$H_{Kerr} = (\hat{x}^2 + \hat{p}^2)^2$,
	may be easier experimentally,
	\item an interaction to couple two modes together, e.g., the
	beam splitter operation.
\end{enumerate}
It is easy to see how this works when the continuous quantity is
the position, $x$ (say).  After evolving the state with some Hamiltonian,
measuring the new position gives us the answer (or one possible answer
out of a superposition or distribution).  But the Heisenberg uncertainty
principle means that if we know the position exactly, the momentum is
totally undetermined.  This asymmetry in the uncertainty is known
as squeezing, and using a single quadrature like this requires infinitely
squeezed states, which are unphysical.  Experimentally, we can't even 
get a good approximation of an infinitely squeezed state, so for
practical purposes, we need to choose a different way to encode our
information, and the obvious choice is to use Gaussian states, which
have the uncertainty spread between the two quadratures $x$ and $p$.
The nonlinear operation in our CVQC will evolve the state into
something that is no longer a Gaussian state: if it didn't, the
whole computation would be efficiently classically simulatable
\citep{bartlett02a}.  This corresponds to having a superposition or
distribution of different answers: when we make a measurement, we
will interpret the outcome as a Gaussian state from which we can
extract the result of out computation.

Implementing CVQC in a real system may require some modification to the
set of elementary gates, to adapt it to what is experimentally practical,
for an example, see \cite{wagner09a}.  
While continuous variables are a practical choice for many quantum
communications tasks (see \cite{braunstein04a} for a review),
very little has been developed for CVQC.  In particular, it shares
the unfavorable scaling of resources with precision that
classical analogue and quantum simulation have, and standard
error correction results for digital quantum computation don't apply
\citep{niset08a}.

\section{CVQC for quantum simulation}
\label{sec:cvqsim}

For quantum simulation, where the digital advantage of binary 
encoding is already lost, CVQC deserves a critical second look. 
Many quantum systems are naturally continuous:
for these a CVQC quantum simulation would be especially appropriate.
Experimental systems suitable for CVQC can be developed from many
of the architectures currently proposed for digital quantum computing:
whenever qubits are coupled using some sort of field mode,
this can be inverted by regarding the field mode as the continuous
variable and the qubits as the control system.
\cite{wagner09a} describes how the micromaser can be used
this way, and the photonic module of \cite{devitt07a} is also
promising.  Squeezing is one of the elementary operations
we need in CVQC when encoding with Gaussian states,
and we can use the degree of squeezing to estimate how large our
system needs to be to beat classical simulation.
\cite{suzuki06a} have achieved 7dB of squeezing in optical experiments.
For one mode, 7dB of squeezing corresponds to $2-3$ bits of precision
i.e., around 5 distinguishable outcomes when the state is measured.
With arbitrary perfect coupling, going beyond what we can simulate
classically corresponds to more than 17 modes coupled together,
since the combination takes us above 40 equivalent (qu)bits.
Experimentally, nine modes have been combined in CV error correction
by \cite{aoki09a}, though this stays within Gaussian states and is
thus still efficiently classically simulatable.
This is a very rough estimate, because evolving the system accurately in 
a classical simulation would in practice require higher precision to
produce the required final precision for the output.  On the other
hand, we can't achieve arbitrary coupled states with current technology.
But it does give a feel for the potential for CVQC quantum simulation.

\section{The future of quantum simulation}
\label{sec:outlook}

The goal of building a quantum computer large enough to
solve problems beyond the reach of the classical computational power
available to us, is coming closer.
The first useful application for quantum computers is likely to be
simulation of quantum systems.  Small quantum simulations have
already been demonstrated, and useful quantum simulations -- that
solve problems inaccessible to classical computation -- require
less than a hundred qubits and may be viable without error correction.
While the theoretical underpinnings of digital computation are well 
developed, there are significant gaps in our understanding of analogue 
computing theory, and these are in turn slowing the progress of CVQC and 
quantum simulation.  Investment in developing the missing 
theory is as crucial for the future development of quantum technology as 
experimental progress in quantum coherence and quantum control.

\begin{acknowledgements}
We thank
Katherine Brown,
Simon Devitt,
Mark Everitt,
Martin Jones,
and
Rob Wagner
for useful and stimulating discussions.
Thanks to Mark Everitt and Rob Wagner for drafts of figure 3.
VK is funded by a Royal Society University Research Fellowship.
WM and KN acknowledge funding from the EU project HIP.
This work was supported by a UK Royal Society International Joint
Project Grant.
\end{acknowledgements}

\small


\label{lastpage}
\end{document}